%% file: sgr_proceeding.tex
\documentclass{PoS}

\usepackage{siunitx}
\usepackage[]{units}

\title{Sgr A* Observations with H.E.S.S. II}

\ShortTitle{Sgr A* Observations with H.E.S.S. II}

\author{\speaker{R.D. Parsons}\\
       Max-Planck-Institut f{\"u}r Kernphysik, Heidelberg, Germany\\
        E-mail: \email{daniel.parsons@mpi-hd.mpg.de}}
\author{{M. Holler}\\
Laboratoire Leprince-Ringuet, Palaiseau, Ecole Polytechnique,
CNRS/IN2P3, France  \\
        E-mail: \email{holler@llr.in2p3.fr}}
\author{{J. King}\\
       Max-Planck-Institut f{\"u}r Kernphysik, Heidelberg, Germany\\
        E-mail: \email{johannes.king@mpi-hd.mpg.de}}
\author{{V. Lefranc}\\
DSM/Irfu, CEA Saclay, Gif-Sur-Yvette Cedex, France\\
        E-mail: \email{valentin.lefranc@cea.fr}}
\author{{E. Moulin}\\
DSM/Irfu, CEA Saclay, Gif-Sur-Yvette Cedex, France\\
        E-mail: \email{emmanuel.moulin@cea.fr}}
\author{{H. Poon}\\
       Max-Planck-Institut f{\"u}r Kernphysik, Heidelberg, Germany\\
        E-mail: \email{cxinio@mpi-hd.mpg.de}}
\author{{ J. Veh}\\
  Universit{\"a}t Erlangen-N{\"u}rnberg, Physikalisches Institut,
  Erwin-Rommel-Str. 1, Erlangen, Germany\\
  E-mail: \email{johannes.veh@physik.uni-erlangen.de}}
\author{{A. Viana}\\
       Max-Planck-Institut f{\"u}r Kernphysik, Heidelberg, Germany\\
        E-mail: \email{aion.viana@mpi-hd.mpg.de}}

\author{{on behalf of the H.E.S.S. Collaboration}}

\abstract{The Galactic Centre has been studied with the High
        Energy Stereoscopic System
        (H.E.S.S.) for over 10 years, revealing a bright, complex
        gamma-ray morphology. Besides a strong
        point-like very-high-energy gamma-ray source coincident with
        the supermassive black hole Sgr A*, previous analyses also
        revealed a diffuse ridge of gamma-ray emission, indicative of
        a powerful cosmic-ray accelerator in this region.

        The addition of a fifth telescope with 600 m$^2$ mirror area to
        the centre of the H.E.S.S. array has increased
        the energy range accessible, allowing observations to take
        place down to 100 GeV and potentially below. This wider energy range allows an
        important overlap in observations with satellite instruments
        such as the Fermi-LAT gamma-ray telescope. We will present the
        results of new H.E.S.S observations of the Galactic Centre
        region and show a detailed analysis of the central source,
        including comparisons to results at other wavelengths.}

\FullConference{The 34th International Cosmic Ray Conference,\\
		30 July- 6 August, 2015\\
		The Hague, The Netherlands}

\begin{document}

\input{sections}

\bibliographystyle{aa}

\input{bibitem.bbl}
\end{document}

%% file: sections.tex
\section{Introduction}

The gamma-ray emission from the Galactic Centre (GC) was first detected by the EGRET satellite in the energy band 100 MeV to 10 GeV (3EG J1746-2852)~\cite{1998AA335161M}. Observations with imaging atmospheric Cherenkov telescopes soon after gave rise to the detection of an emission in the very-high-energy (VHE, E $\gtrsim$ 100 GeV) gamma-ray regime \cite{Tsuchiya:2004wv, Kosack:2004ri, Aharonian:2004wa, Albert:2005kh}. The Fermi-LAT satellite has detected a source at the GC in the energy range between 20 MeV to more than 300 GeV \cite{2009ApJS..183...46A}, however due to the strong gamma-ray diffuse background in this energy range no conclusion was taken about possible associations of the Fermi-LAT source with other gamma-ray sources.

The observation of the GC region with the H.E.S.S.~instrument led to the detection of a point-like source of VHE gamma-rays, HESS J1745-290, with an unprecedented accuracy in the TeV energy range. While spatially coincident with the supermassive black hole Sgr A*, the position of HESS J1745-290 was still compatible with the supernova remnant Sgr A East, and the plerion G359.95-0.04, despite the angular resolution of the H.E.S.S. instrument of about 6$'$. After a careful investigation of the pointing systematics of the H.E.S.S. telescopes, the systematic error on the centroid position of HESS J1745-290 emission was reduced to 13{\rm$''$}, allowing the exclusion of Sgr A East as the main counterpart of the emission~\cite{2010MNRAS.402.1877A}. A larger exposure on the GC region revealed a ridge of diffuse emission extending along the Galactic plane for about 2$^{\circ}$ in Galactic longitude, which was found to be spatially correlated to giant molecular clouds located in the central molecular zone~\cite{Aharonian:2006au}. The strong correlation between the morphology of the diffuse gamma-ray emission and the density of molecular clouds indicates the presence of a proton accelerator in the GC region, since energetic proton interactions with the cloud material would give rise to the observed gamma-ray flux via $\pi^0$ decays. 

The very nature of the VHE central emission remains still unknown, leaving Sgr A*~\cite{Aharonian:2004jr,Liu:2006bf}, G359.95-0.04~\cite{Wang:2005ya,Hinton:2006zk} and a spike of annihilating dark matter~\cite{2012PhRvD..86h3516B} as possible counterparts for the observed emission. The H.E.S.S. experiment has been taking observations of the GC region for almost ten years with the full 4-telescope array during its first phase of operation and the data collected by H.E.S.S.~I allows for the most detailed high energy gamma-ray picture reported to date of the GC region. In 2012, the addition of a fifth telescope, CT\,5, with 600m$^2$ mirror area to the centre of the H.E.S.S. array has marked the start the second phase of the H.E.S.S. experiment. With H.E.S.S~II the accessible energy range is increased, allowing observations to take place down to almost 100 GeV in the GC region. Here we present the analysis of the central source with data taken in the first year of H.E.S.S.~II observations.

\section{H.E.S.S. II Observations}


Observations of the GC region were made with the full 5-telescope H.E.S.S. II array throughout 2013 and early 2014. The analysis of the central source presented here makes use of 132 observation runs taken towards the GC, covering a zenith angle range of 5-45$^\circ$ with a mean zenith angle of 22$^\circ$. The dataset amounts to 58.6 live hours of observations at the nominal position of Sgr A*. All selected runs meet the H.E.S.S standard quality criteria and CT\,5 was required to have been active during the run. In order to reduce the systematic error of the cameras, the data were taken in \textit{wobble mode}, where the pointing direction is chosen at alternating offsets from the target position. As not all observations were made specifically for the central source, the wobble offsets range from 0-1.5$^\circ$, with the majority of offsets lying in the 0.4-0.9$^\circ$ range. 

After calibration of the raw data from photomultiplier tube signals, events are reconstructed using a technique based on a semi-analytical shower development model \cite{model, modelMono}. When analysing H.E.S.S. II data, where two different telescope types participate in one run, three analysis modes are possible: The \textit{mono} analysis uses only data from CT\,5, ignoring any images seen by one of the other telescopes, the \textit{stereo} analysis uses only events where two or more simultaneous shower images are available, and the \textit{combined} analysis chooses the best of the above reconstructions for a given event. The results presented here are based on the \textit{combined} analysis as it is able to provide the low energy performance of the \textit{mono} event reconstruction, with highly accurate reconstruction of stereo events at higher energies.

As the GC is a bright region when observed at optical wavelengths, it contains large variations in the night sky background photon detection rate across the field of view (100-300\,MHz). Particularly in the \textit{mono} analysis, such night sky background triggers can be mistaken for gamma-ray events leading to potential spurious sources and gradients in the field of view. In order to control this effect, additional event selection cuts were placed on the dataset (described in detail in \cite{modelMono}).

In order to ensure the stability of the results presented here, all analyses were cross-checked using an independent calibration and analysis chain \cite{ImPACT}, which gives consistent results.

\subsection{Morphology}

\begin{figure*}
    \centering
\includegraphics[width=0.9\textwidth]{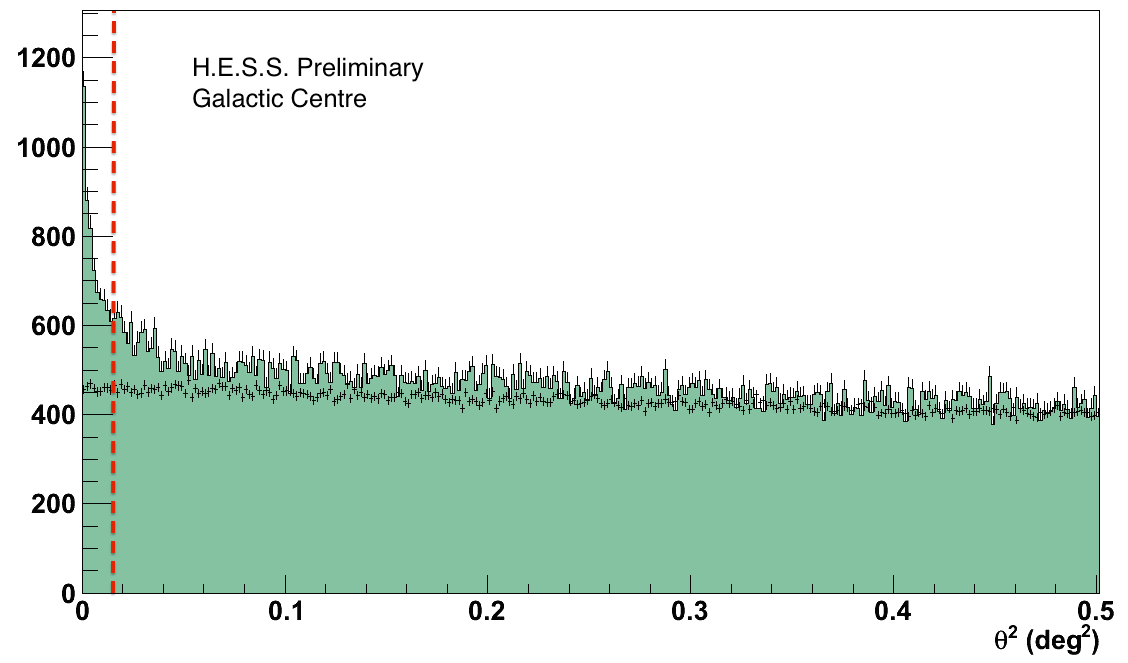}
\caption{Event distribution in the source region (green shaded) compared the expected level of background contamination (black points) at a function of squared angular distance from the position of Sgr A*, created using the \textit{combined} event analysis mode. The vertical red line shows the region used for spectral extraction.}
	\label{fig-thetasqr}
\end{figure*}

Figure \ref{fig-thetasqr} shows the event distribution as a function of squared angular distance from the position of Sgr A*, comparing that seen in the source region with the estimated level of background contamination using the \textit{reflected regions} method \cite{berge2006}. In this figure an excess source is clearly visible at the position of Sgr A* at a significance of 40$\sigma$ with around 3600 excess events in the central 0.015 deg$^2$ and a relatively long tail stretching out to 0.3 deg$^2$. Such a tail quite clearly shows the contribution of diffuse gamma-ray emission at large distances from the source.

\begin{figure*}
    \centering
\includegraphics[width=0.49\textwidth]{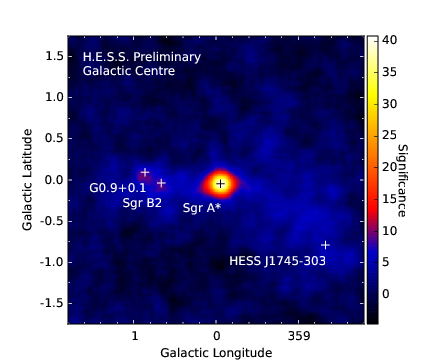}
\includegraphics[width=0.49\textwidth]{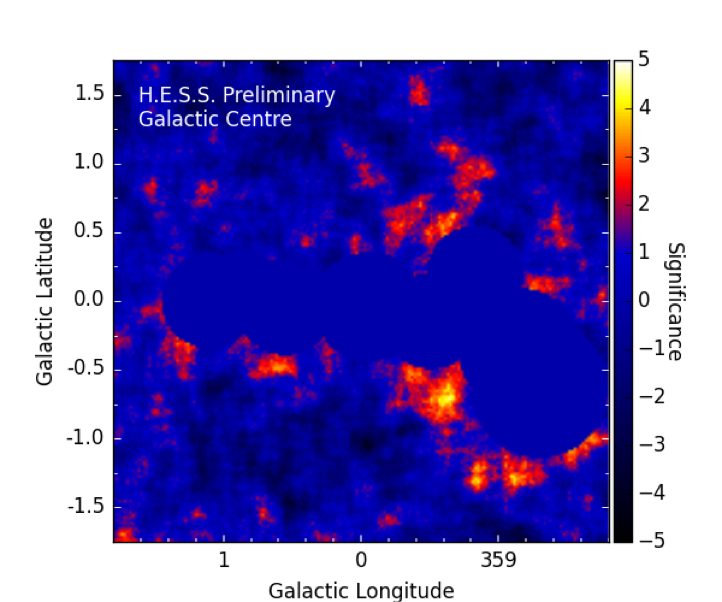}

\caption{Significance map of the GC region, created using the \textit{combined} event analysis mode (left). Significance distribution of events falling outside of excluded regions, i.e. regions where no strong gamma-ray emission is expected. Events are oversampled with a radius of 0.1$^\circ$. }
	\label{fig-map}
\end{figure*}

Figure \ref{fig-map} shows the gamma-ray significance map in the region around the central source.  The estimated level of background contamination in the map is determined using the \textit{ring} method \cite{berge2006}, using a ring size adaptively scaled to allow a good background estimation across the full field. Using this method the central source is clearly seen at a position compatible with previous H.E.S.S. analyses. In addition to the central source, gamma-ray emission is seen at the position of the previously detected pulsar wind nebula G0.9+0.1. Between these sources some diffuse gamma-ray emission is seen along the Galactic plane, however due to the relatively limited angular resolution in the low-energy regime ($R_{68\%} \approx 0.15-0.2^{\circ}$) the structure of this emission is difficult to determine. Figure \ref{fig-map}  also shows the significance distribution of the events falling outside the regions of expected gamma-ray emission. This significance map is consistent with no additional emission in the GC region, except in some areas bordering the currently excluded region, where some additional features are seen at the 3-4 $\sigma$ level. These features may be an indication of either some extension of the emission previously seen by H.E.S.S., or escape of emission outside the excluded region due to the reduced angular resolution compared to previous observations.

\begin{figure*}
  \includegraphics[width=0.9\textwidth]{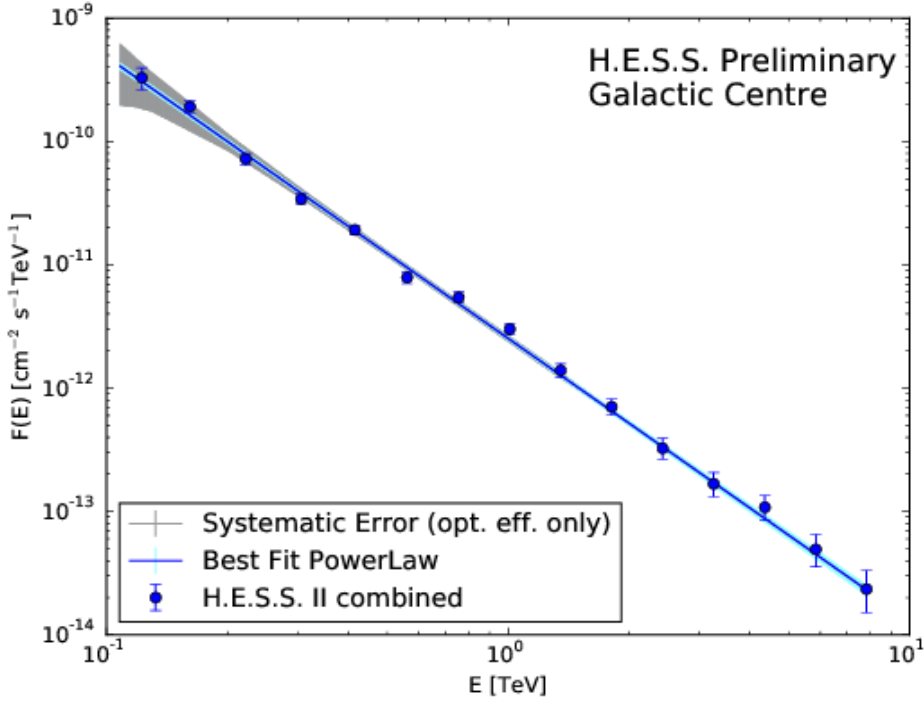}
    \centering
\caption{Spectral results of the point-source analysis of the Sgr A* region, based on integration of the events within  0.12$^\circ$ of the source position. The power-law spectral fit is shown, with its one sigma statistical error band in blue, grey shaded area shows estimated systematic uncertainty from differing optical efficiencies in the H.E.S.S. telescopes. }
	\label{fig-spectral}
\end{figure*}

\newpage
\subsection{Spectrum}

A spectrum was extracted from a circular region with radius 0.12$^\circ$ around the central source as indicated in figure \ref{fig-thetasqr}. In this case the level of background contamination in the source region was estimated using the \textit{reflected regions} method \cite{berge2006}. In the GC region a large number of region has to be excluded from the background estimation because they contain strong gamma-ray sources and for some runs it was not possible to find suitable regions at all. This leads to a reduced live time of only 48.6 hours for the spectral analysis. The spectrum was fit with a power-law model of the form $\phi_0 \left( \frac{E}{E_0} \right)^{-\Gamma}$, using a forward folding approach to fold the spectral model with the expected instrument response. The result of this fit is shown in figure \ref{fig-spectral}. The power-law model provides an acceptable fit to the data and yields a spectrum covering energies from 110\,GeV to almost 10\,TeV, with a power-law index $\Gamma$ of $2.28\pm0.04$ and a normalization $\phi_0 = 9.46\pm0.31 \times 10^{-12}$ cm$^{-2}$s$^{-1}$TeV$^{-1}$ at $E_0$ = 0.562\,GeV. The resulting differential flux at 1\,TeV ($\phi(1\,\mathrm{TeV}) = 2.54\pm0.1 \times 10^{-12}$\,cm$^{-2}$s$^{-1}$TeV$^{-1}$) and integral flux above 1\,TeV ($\phi(> 1\,\mathrm{TeV}) = 1.98\pm0.01 \times 10^{-12}$\,cm$^{-2}$s$^{-1}$) are consistent with previously published H.E.S.S. results \cite{HESSIGC}. For this analysis a conservative systematic error band has been added around the best fit model estimating the uncertainty on the behaviour of the analysis energy threshold due to large differences in the optical efficiency of CT5 in comparison to the 4 smaller telescopes in the array (85\% vs 65\% of the nominal reflectivity).
Although previous H.E.S.S. publications have shown a significant cut-off in the spectrum of this source, no significant evidence for a cut-off is seen in this dataset. This is due to the relatively small exposure of the present analysis in comparison with the much larger dataset available for H.E.S.S. I analysis (93 hours).


\section{Multiwavelength Comparisons}

\begin{figure*}
\includegraphics[width=0.95\textwidth]{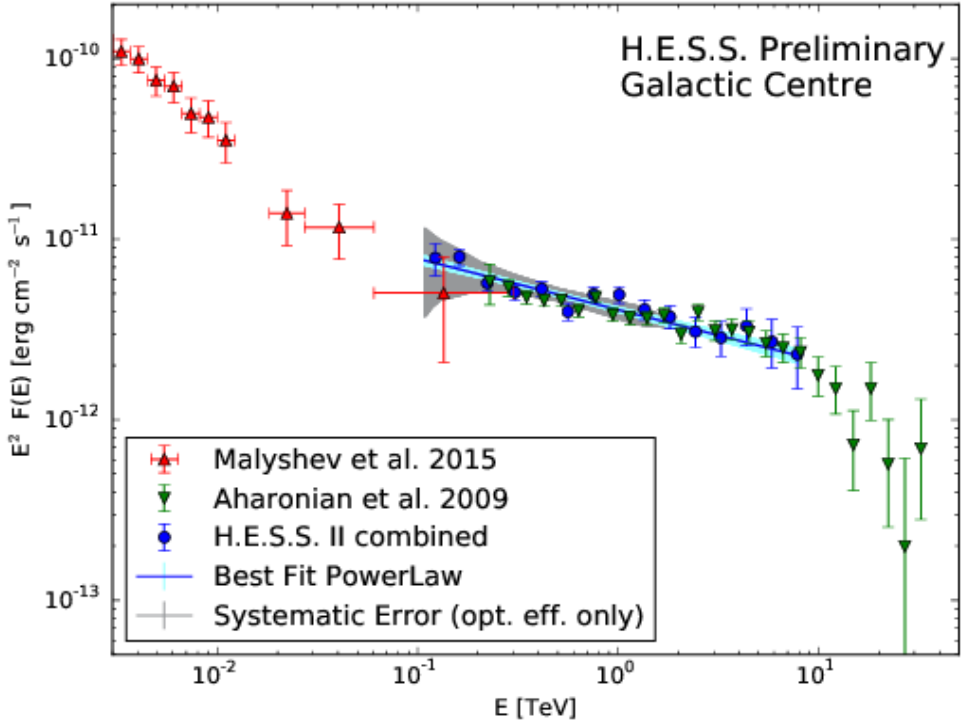}
    \centering
\caption{Spectral energy distribution of the H.E.S.S. II Sgr A* measurements shown in comparison to H.E.S.S. I  \cite{HESSIGC} and Fermi-LAT  \cite{Chernyakova} measurements.}
	\label{fig-SED}
\end{figure*}

Figure \ref{fig-SED} shows a comparison of the H.E.S.S. II spectral energy distribution of the central source to results obtained with H.E.S.S. I \cite{HESSIGC} and the Fermi-LAT \cite{Chernyakova}. 
At low energies, the H.E.S.S. II observations extend down to 110 GeV, into the energy range covered by the Fermi-LAT telescope. Comparison of the H.E.S.S II flux level with that of the Fermi-LAT in the overlapping energy range shows good consistency. However, it should be noted that the H.E.S.S. spectrum was derived using aperture photometry, i.e. not taking into account any possible contribution of diffuse background to the spectrum. In contrast, the Fermi spectrum contains some modelling of diffuse emission which may cause differences in the derived spectrum.  Comparison of the spectral indices in the Fermi and H.E.S.S.  shows clear evidence of a break in the gamma-ray spectrum between 20 and 200 GeV. However, the low event statistics of both the Fermi and H.E.S.S. observations in this crossover region make the nature of this break difficult to constrain.

\section{Conclusions}

We have presented a new analysis of the Galactic Centre region using the full array of 5 H.E.S.S. telescopes. In doing so we have expanded the energy range of spectral observations into a regime previously unexplored by ground based observations (110\,GeV). Comparison to Fermi-LAT and previous H.E.S.S. spectral measurements show an excellent match in the overall flux and spectral shape. Unfortunately, with the current energy threshold it has not been possible to measure the exact shape of the spectral break between the Fermi and H.E.S.S. observations. However, further H.E.S.S. II observations have already been made of this region, with an increased understanding of systematic uncertainties which should allow a significantly improved low energy analysis of this region in the near future.

\vspace*{0.5cm}
\footnotesize{{\bf Acknowledgment:}{The support of the Namibian authorities and of the University of Namibia in facilitating the construction and operation of H.E.S.S. is gratefully acknowledged, as is the support by the German Ministry for Education and Research (BMBF), the Max Planck Society, the German Research Foundation (DFG), the French Ministry for Research, the CNRS-IN2P3 and the Astroparticle Interdisciplinary Programme of the CNRS, the U.K. Science and Technology Facilities Council (STFC), the IPNP of the Charles University, the Czech Science Foundation, the Polish Ministry of Science and Higher Education, the South African Department of Science and Technology and National Research Foundation, and by the University of Namibia. We appreciate the excellent work of the technical support staff in Berlin, Durham, Hamburg, Heidelberg, Palaiseau, Paris, Saclay, and in Namibia in the construction and operation of the equipment.}}

%% file: sgr_proceeding.bbl
\begin{thebibliography}{13}
\expandafter\ifx\csname natexlab\endcsname\relax\def\natexlab#1{#1}\fi


\bibitem[1]{1998AA335161M}
{Mayer-Hasselwander}, H.~A. {et~al.} 1998, A\&A, 335, 161

\bibitem[2]{Tsuchiya:2004wv}
Tsuchiya, K. {et~al.} 2004, Astrophys.J., 606, L115

\bibitem[3]{Kosack:2004ri}
Kosack, K. {et~al.} 2004, Astrophys.J., 608, L97

\bibitem[4]{Aharonian:2004wa}
Aharonian, F. {et~al.} 2004, Astron.Astrophys., 425, L13

\bibitem[5]{Albert:2005kh}
Albert, J. {et~al.} 2006, Astrophys.J., 638, L101

\bibitem[6]{2009ApJS..183...46A}
{Abdo}, A.~A. {et~al.} 2009, Astrophysical Journal, Supplement, 183, 46

\bibitem[7]{2010MNRAS.402.1877A}
{Acero}, F., others, \& {H.E.S.S.~Collaboration}. 2010, Mon. Not. Roy. Ast.
  Soc., 402, 1877

\bibitem[8]{Aharonian:2006au}
Aharonian, F. {et~al.} 2006, Nature, 439, 695

\bibitem[9]{Aharonian:2004jr}
Aharonian, F. \& Neronov, A. 2005, Astrophys.J., 619, 306

\bibitem[10]{Liu:2006bf}
Liu, S.-M., Melia, F., Petrosian, V., \& Fatuzzo, M. 2006, Astrophys.J., 647,
  1099

\bibitem[11]{Wang:2005ya}
Wang, Q.~D., Lu, F., \& Gotthelf, E. 2006, Mon.Not.Roy.Astron.Soc.,
367, 937

\bibitem[12]{Hinton:2006zk}
Hinton, J. \& Aharonian, F. 2007, Astrophys.J., 657, 302

\bibitem[13]{2012PhRvD..86h3516B}
{Belikov}, A.~V., {Zaharijas}, G., \& {Silk}, J. 2012, Physical Review D, 86, 083516

\bibitem[14]{model}
de Naurois, M. \& Rolland, L. 2009, Astroparticle Physics,
  32:231-252, 2009.

\bibitem[15]{modelMono}
Holler, M., Berge, D., van Eldik, C. 2015, 
Proceedings of the 34th International Cosmic Ray Conference, 1046

\bibitem[16]{ImPACT}
Parsons, R.D., Gajdus, M. \& Murach, T. 2015, 
Proceedings of the 34th International Cosmic Ray Conference, 928


\bibitem[17]{berge2006}
{Berge}, D., {Funk}, S. and {Hinton}, J. 2007, Astron.Astrophys., 466, 1219

\bibitem[18]{HESSIGC}
Aharonian, F. {et~al.} 2009, Astron.Astrophys., 503, 817

\bibitem[19]{Chernyakova}
{Malyshev}, D., {Chernyakova}, M., {Neronov}, A., {Walter}, R. 2015, arXiv:1503.05120

\end{thebibliography}
